\documentclass[11pt,a4paper,twoside]{article}
\usepackage{graphicx}
\usepackage{color}
\begin{document}
\baselineskip=0.20in
\vspace{20mm}
\baselineskip=0.30in
\begin{center}

{\large \bf \Large Bound state solutions of the Manning–Rosen potential}

\vspace{4mm}

{ Falaye, B.\ J.\footnote{fbjames11@physicist.net}, Oyewumi, K.\,J.\footnote{kjoyewumi66@unilorin.edu.ng}, Ibrahim, T. T. \footnote{ibrahimng@yahoo.com}, Punyasena, M. A. \footnote{ Department of Physics, University of Kelaniya, Kelaniya, Sri Lanka: sena@kln.ac.lk} and Onate, C. A. \footnote{onateca12@gmx.us}} 

{Theoretical Physics Section, Department of Physics\\ University of Ilorin,  P. M. B. 1515, Ilorin, Nigeria. }

\vspace{4mm}

\end{center}

\noindent
\begin{abstract}
\noindent
Using the asymptotic iteration method (AIM), we have obtained analytical approximations to the $\ell$-wave solutions of the Schr\"{o}dinger equation with the Manning-Rosen potential. The equation of energy eigenvalues equation and the corresponding wavefunctions have been obtained explicitly. Three different Pekeris-type approximation schemes have been used to deal with the centrifugal term. To show the accuracy of our results, we have calculated the eigenvalues numerically for arbitrary quantum numbers $n$ and $\ell$ for some  diatomic molecules (HCl, CH, LiH and CO). It is found that the results are in good agreement with other results found in the literature. A straightforward extension to the s-wave case and Hulth$\acute{e}$n potential case are also presented.
\end{abstract}


{\bf PACs No.} 03.65.Ge; 03.65.-w;02.30.Gp

\section{Introduction}
In recent years, the problem of analytical solution of Schr\"{o}dinger equation with angular momentum quantum number $\ell=0$ and $\ell \neq 0$ for a number of exponential-type potentials has been addressed by many researchers. Some of these exponential-type potentials include Manning-Rosen potential \cite{WCQ07, SHD07, ADA05, WCQ09, MFM33}, Morse potential \cite{CBJ05, JPK02}, Hulth$\acute{e}$n potential \cite{OBI07, MAR04, LHA42}, Woods-Saxon potential \cite{BC06}, P\"{o}schl-Teller potential \cite{KJ10}, Rosen-Morse potential \cite{KJ10B}, Deng-Fan potential \cite{DG11} and trigonometric Scarf potential \cite{BJ10}. 

It is known that radial Schr\"{o}dinger equation for these potentials can be solved exactly for $\ell=0$. On the other hand, it is also known that for these potentials, the Schr\"{o}dinger equation cannot be solved for $\ell\neq0$. To obtain the solution for $\ell\neq0$, one has to use Pekeris-type approximation scheme to deal with the centrifugal term or solve numerically. The most widely used approximation was introduced by Pekeris \cite{CLP34} and another form was suggested by Greene et al. \cite{RLG76} and Qiang et al. \cite{WCQ09}.

The approximate analytical solution of the Schr\"{o}dinger equation with Manning-Rosen potential has been studied by many researchers \cite{WCQ07, WCQ09, SMI11, WLF99}. This potential has been used in several branches of physics for bound states and scattering properties. The Dirac equation with this potential has been investigated also, the approximate  solutions of the effective mass for this potential in $N$ dimensions has been obtained by using the asymptotic iteration method \cite{BaY12}. Recently, Hassanabadi et al. (2012) used SUSY approach to obtain the spin pseudospin symmetries of the Dirac equation with the actual and general Manning-Rosen potentials \cite{HaE12}.

In addition, it also gives an excellent description of the interaction between two atoms in diatomic molecules. The short range Manning-Rosen potential is given by \cite{WCQ07, ADA05, WCQ09, MFM33}
\begin{equation}
V(r)=\frac{-A\hbar^2}{2\mu b^2}\frac{e^{-r/b}}{1-e^{-r/b}}+\frac{\alpha(\alpha-1)\hbar^2}{2\mu b^2}\left[\frac{e^{-r/b}}{1-e^{-r/b}}\right]^2
\label{E1},
\end{equation}
where $A$ and $\alpha$ are dimensioness parameters, while $b$ is the screening parameter which has dimension of length.

Motivated by the success in obtaining analytical solution of the Schr\"{o}dinger equation with potential equation (\ref{E1}) using the Nkiforov-Uvarov method (N-U) \cite{SKN88} by Ikhdair \cite{SMI11}, the standard method by Qiang et al. \cite{WCQ07, WCQ09} and the numerical integration procedure by Lucha and Sh\"{o}berl \cite{WLF99}. We attempt to use a different and more practical method called the asymptotic iteration method (AIM) \cite{HCR03, HCR05, Fal12} within the framework of the Pekeris approximations suggested by Qiang et al. \cite{WCQ09} to solve the Schr\"{o}dinger equation with short range Manning-Rosen potential for any arbitrary $\ell$-state. These approximations are \cite{WCQ09} :
\begin{equation}
Approx. 1:~~~~   \frac{1}{r^2}\approx\frac{1}{b^2}\left[\frac{1}{12}+\frac{e^{-r/b}}{(1-e^{-r/b})^2}\right],
\label{E2}
\end{equation}
\begin{equation}
Approx. 2:~~~~~    \frac{1}{r^2}\approx\frac{1}{b^2}\left[\frac{e^{(1-r)/b}}{1-e^{-r/b}}+\frac{e^{-2r/b}}{(1-e^{-r/b})^2}\right].
\label{E3}
\end{equation}
These approximations are proposed instead of the commonly used one by others \cite{WCQ09}:
\begin{equation}
Approx. 3:~~~~~ \frac{1}{r^2}\approx\frac{1}{b^2}\frac{e^{-r/b}}{(1-e^{-r/b})^2}.
\label{E4}
\end{equation}
Equations (\ref{E2}) and (\ref{E3}) are more general than equation (\ref{E4}), while equation (\ref{E2}) and (\ref{E3}) give a better approximation to the centrifugal term when $b$ is small \cite{WCQ09}.
This paper is organized as follows. In section $2$, the review of the asymptotic iteration method (AIM) is presented. In section $3$, this method is applied to radial Schr\"{o}dinger equation to find the analytical solution. The numerical calculations are given and the results are compared with those obtained by other methods \cite{WCQ09,SMI11, WLF99} in section $4$. In section $5$, a brief conclusion is given. 

\section{The Asymptotic Iteration Method (AIM)}
The Asymptotic Iteration Method (AIM) is proposed to solve the homogenous linear second-order differential equation of the form
\begin{equation}
y_n''(x)=\lambda_o(x)y_n'(x)+s_o(x)y_n(x),
\label{E5}
\end{equation}
where $\lambda_o(x)\neq0$ and the prime denote the derivative with respect to $x$, the extra parameter $n$ can be described as a radial quantum number $\cite{HCR03, HCR05}$. The variables, $s_o(x)$ and  $\lambda_o(x)$ are sufficiently differentiable. To find a general solution to this equation, we differentiate equation (\ref{E5}) with respect to $x$, we find
\begin{equation}
y_n'''(x)=\lambda_1(x)y_n'(x)+s_1(x)y_n(x),
\label{E6}
\end{equation}
where
\begin{eqnarray}
\lambda_1(x)&=&\lambda_o'(x)+s_o(x)+\lambda_o^2(x),\nonumber\\
s_1(x)&=&s_o'(x)+s_o(x)\lambda_o(x).
\label{E7}
\end{eqnarray}
The second derivative of equation (\ref{E5}) is obtained as
\begin{equation}
y_n''''(x)=\lambda_2(x)y_n'(x)+s_2(x)y_n(x),
\label{E8}
\end{equation}
where
\begin{eqnarray}		
\lambda_2(x)&=&\lambda_1'(x)+s_1(x)+\lambda_o(x)\lambda_1(x),\nonumber\\
	 s_2(x)&=&s_1'(x)+s_o(x)\lambda_1(x).
\label{E9}
\end{eqnarray}
Equation (\ref{E5}) can be iterated up to $(k+1)^{th}$ and $(n+2)^{th}$ derivatives, $k=1,2,3, \ldots$, therefore, we have
\begin{eqnarray}
y_n^{(k+1)}(x)&=&\lambda_{k-1}(x)y_n'(x)+s_{k-1}(x)y_n(x),\nonumber\\
y_n^{(k+2)}(x)&=&\lambda_{k}(x)y_n'(x)+s_{k}(x)y_n(x),
\label{E10}
\end{eqnarray}
where
\begin{eqnarray}
\lambda_k(x)&=&\lambda_{k-1}'(x)+s_{k-1}(x)+\lambda_o(x)\lambda_{k-1}(x),\nonumber\\
s_k(x)& = &s_{k-1}'(x)+s_{o}(x)\lambda_{k-1}(x).
\label{E11}
\end{eqnarray}
From the ratio of the $(k+2)^{th}$ and $(k+1)^{th}$ derivatives
\begin{equation}
\frac{d}{dx} \ln \left[y_n^{k+1}(x)\right]=\frac{y_n^{(k+2)}(x)}{y_n^{(k+1)}(x)}=\frac{\lambda_k(x)\left[y_n'(x)+\frac{s_k(x)}{\lambda_k(x)}y_n(x)\right]}{\lambda_{k-1}(x)\left[y_n'(x)+\frac{s_{k-1}(x)}{\lambda_{k-1}(x)}y_n(x)\right]},
\label{E12}
\end{equation}
if $k>0$, for sufficiently large $k$, we obtain $\alpha$ values from
\begin{equation}
\frac{s_k(x)}{\lambda_k(x)}=\frac{s_{k-1}(x)}{\lambda_{k-1}(x)}=\alpha(x),
\label{E13}
\end{equation}
with quantization condition
\begin{equation}
\delta_k(x)=
\left|
\begin{array}{lr}     
\lambda_k(x)&s_k(x) \\      
  \lambda_{k-1}(x)&s_{k-1}(x)
  \end{array}
  \right|=0\ \ ,\ \  \ k=1, 2, 3....
\label{E14}
  \end{equation}
Then, equation (\ref{E12}) reduces to 
\begin{equation}
\frac{d}{dx} \ln \left[y_n^{(k+1)}(x)\right]=\frac{\lambda_k(x)}{\lambda_{k-1}(x)},
\label{E15}
\end{equation}
which yields the general solution of equation (\ref{E5}) as
\begin{equation} y(x)=\exp\left(-\int^x\alpha(x')dx'\right)\left[C_2+C_1\int^x\exp\left(\int^{x'}\left[\lambda_o(x'')+2\alpha(x'')\right]dx''\right)dx'\right].
\label{E16}
\end{equation}
For a given potential, the idea is to convert the radial Schr$\ddot{o}$dinger equation to the form of equation (\ref{E5}). Then, $\lambda_o(x)$ and $s_o(x)$ are determined and $s_k(x)$ and $\lambda_k(x)$ parameters are obtained by the recurrence relations given by equation (\ref{E11}). The energy eigenvalues are then obtained by the condition given by equation (\ref{E14}) if the problem is exactly solvable.

Suppose we wish to solve the radial Schr$\ddot{o}$dinger equation for which the homogeneous linear second-order differential equation takes the following general form
\begin{equation}
y_{n}''(x)=2\left(\frac{ax^{N+1}}{1-bx^{N+2}}-\frac{m+1}{x}\right)y_{n}'(x)-\frac{Wx^N}{1-bx^{N+2}}.
\label{E17}
\end{equation}
The exact solution $y_n(x)$ can be expressed as \cite{HCR05}
\begin{equation}
y_n(x)=(-1)^nC_2(N+2)^n(\sigma)_{_n}{_2F_1(-n,\rho + n;\sigma; bx^{N+2})}
\label{E18},
\end{equation}
where the following notations have been used:
\begin{equation}
(\sigma)_{_n}=\frac{\Gamma{(\sigma+n)}}{\Gamma{(\sigma)}}\ \ ,\ \ \sigma=\frac{2m+N+3}{N+2}\  \mbox{and}\ \ \rho=\frac{(2m+1)b+2a}{(N+2)b}.
\label{E19}
\end{equation}

\section{The Eigenvalues and Eigenfunctions of the Schr\"{o}dinger Equation with the Manning-Rosen Potential}
The Schr\"{o}dinger equation is given by
\begin{equation}
\left\{-\frac{\hbar^2}{2\mu}\left[\frac{1}{r^2}\frac{\partial}{\partial r}r^2\frac{\partial}{\partial r}+\frac{1}{r^2\sin \theta}\frac{\partial}{\partial\theta}\left(\sin\theta\frac{\partial}{\partial\theta}\right)+\frac{1}{r^2\sin^2\theta}\frac{\partial^2}{\partial\phi^2}\right]+V(r)-E\right\}\psi(r)=0.
\label{E20}
\end{equation}
By taking $\psi(r)=R_{n\ell}(r)Y_{\ell m}(\theta,\phi)r^{-1}$ and considering potential (\ref{E1}), we obtain the radial part of the Schr\"{o}dinger equation as:
\begin{equation}
\frac{d^2R_{nl}(r)}{dr^2}+\left[\frac{2\mu E_{n\ell}}{\hbar^2}-V_{eff}\right]R_{n\ell}(r)=0,
\label{E21}
\end{equation}
where
\begin{equation}
V_{eff}(r)=V_{MR}(r)+V_{\ell}(r)=\frac{1}{b^2}\left[\frac{\alpha(\alpha-1)e^{-2r/b}}{(1-e^{-r/b})^2}-\frac{Ae^{-r/b}}{1-e^{-r/b}}\right]+\frac{\ell(\ell+1)}{r^2}.
\label{E22}
\end{equation}
By putting the approximation expression (\ref{E2}) into equation (\ref{E21}) and using $z=(e^{-r/b}-1)^{-1}$, we have
\begin{equation}
z(z+1)\frac{d^2R_{n\ell}(z)}{dz^2}+(1+2z)\frac{dR_{n\ell}(z)}{dz}+\frac{\left(\Lambda_a z^2+\Lambda_bz+\Lambda_c\right)R_{n\ell}(z)}{z+1}=0,
\label{E23}
\end{equation}
where, we have used the following notations for simplicity:
\begin{eqnarray}
\Lambda_b-\Lambda_c-\Lambda_a&=& \frac{\ell(\ell+1)}{12}-\frac{2\mu E_{n\ell}b^2}{\hbar^2},\nonumber\\
\Lambda_b-\Lambda_a&=&-\alpha(\alpha-1)-A,\nonumber\\
-\Lambda_a&=&\alpha(\alpha+1)+\ell(\ell+1).
\label{E24}
\end{eqnarray}
In order to solve (\ref{E23}) with the AIM, we need to transform this equation into the form of equation (\ref{E5}). Therefore, the reasonable physical wave function can be obtained as:
\begin{equation}
R_{n\ell}(z)=z^\gamma(z+1)^{-\beta}f_{n\ell}(z),
\label{E25}
\end{equation}
where $\gamma=\sqrt{-\Lambda_c}$ and $\beta=\sqrt{\Lambda_b-\Lambda_a-\Lambda_c}$.\\
With equation (\ref{E25}), equation (\ref{E23}) turns to a second-order homogeneous linear differential equation of the form
\begin{equation} \frac{d^2f_{n\ell}(z)}{dz^2}=\left\{\frac{2\beta-1}{1+z}-\frac{2\gamma+1}{z}\right\}\frac{df_{n\ell}(z)}{dz}+\left\{\frac{(\beta-\gamma)-(\beta-\gamma)^2-\Lambda_a}{1+z}\right\}f_{n\ell}(z).
\label{E26}
\end{equation}
We now apply the AIM in solving equation (\ref{E26}). Comparing equations (\ref{E5}) and (\ref{E26}), we have the values of $\lambda_n$ and $s_n$ as follows:
\begin{eqnarray}
\lambda_o& = &\left \{ \frac{2\beta-1}{1+z}-\frac{2\gamma+1}{z} \right \}, \nonumber \\ 
s_o& = & \left \{ \frac{(\beta-\gamma)-(\beta-\gamma)^2-\Lambda_a}{1+z} \right \}, \nonumber \\ \lambda_1 & = & \left \{ \frac{1-2\beta}{(z+1)^2}+\frac{2\gamma+1}{z^2}+\frac{(\beta-\gamma-(\beta-\gamma)^2-\Lambda_a)}{1+z}+\left(\frac{2\beta-1}{1+z}-\frac{2\gamma+1}{z}\right)^2\right\}, \nonumber \\ s_1& = & \left \{ \frac{(\gamma-\beta+(\gamma-\beta)^2+\Lambda_a)}{(1+z)^2}+\left(\frac{\beta-\gamma-(\beta-\gamma)^2-\Lambda_a}{1+z}\right)\left(\frac{2\beta-1}{1+z}-\frac{2\gamma+1}{z}\right) \right \}, \nonumber \\
\ldots etc.
\label{E27}
\end{eqnarray}
Combining these results with the condition given by equation (\ref{E14}) yields:
\begin{eqnarray}
\frac{s_o}{\lambda_o}&=&\frac{s_1}{\lambda_1}\ \ \  \Rightarrow \ \ \   \gamma_o-\beta_o=-\frac{1}{2}-\frac{1}{2}\sqrt{1-4\Lambda_a},\nonumber\\
\frac{s_1}{\lambda_1}&=&\frac{s_2}{\lambda_2}\ \ \  \Rightarrow \ \ \   \gamma_1-\beta_1=-\frac{3}{2}-\frac{1}{2}\sqrt{1-4\Lambda_a},\nonumber\\
\frac{s_2}{\lambda_2}&=&\frac{s_3}{\lambda_3}\ \ \  \Rightarrow \ \ \   \gamma_2-\beta_2=-\frac{5}{2}-\frac{1}{2}\sqrt{1-4\Lambda_a},\nonumber\\
\ldots etc.
\label{E28}
\end{eqnarray}
When the above expressions are generalized and using the notations in equation (\ref{E24}), the energy values of the schr\"{o}dinger equation with the Manning-Rosen potential turn out as:
\begin{eqnarray}
 E^{(Approx. 1)}_{n\ell}&=&\left[\alpha(\alpha-1)+A+\frac{\ell(\ell+1)}{12}\right]\frac{\hbar^2}{2\mu b^2}\nonumber\\
&-&\left[\frac{\left[(n+\frac{1}{2})+\frac{1}{2}\sqrt{1+4\alpha(\alpha-1)+4\ell(\ell+1)}\right]^2+\alpha(\alpha-1)+A}{\left[(n+\frac{1}{2})+\frac{1}{2}\sqrt{1+4\alpha(\alpha-1)+4\ell(\ell+1)}\right]}\right]^2\frac{\hbar^2}{8\mu b^2}.
\label{E29}
\end{eqnarray}
Again by using approximation (\ref{E3}) and repeat the above procedure, we can consequently obtain the energy eigenvalues as
\begin{eqnarray}
 E^{(Approx.2)}_{n\ell}=\frac{\zeta(\ell)\hbar^2}{2\mu b^2} -\left[\frac{\left[\left(\frac{1}{2}+n\right)+\frac{1}{2}\sqrt{1+4\alpha(\alpha-1)+4\ell(\ell+1)}\right]^2+\zeta(\ell)}{\left[\left(\frac{1}{2}+n\right)+\frac{1}{2}\sqrt{1+4\alpha(\alpha-1)+4\ell(\ell+1)}\right]}\right]^2\frac{\hbar^2}{8\mu b^2}.
\label{E30}
\end{eqnarray}
Similarly, by using the usual approximation in equation (\ref{E4}), the energy eigenvalues are obtained as
\begin{eqnarray}
 E^{(Approx. 3)}_{n\ell}&=&\left[\alpha(\alpha-1)+A\right]\frac{\hbar^2}{2\mu b^2}\nonumber\\
&-&\left[\frac{\left[(n+\frac{1}{2})+\frac{1}{2}\sqrt{1+4\alpha(\alpha-1)+4\ell(\ell+1)}\right]^2+\alpha(\alpha-1)+A}{\left[(n+\frac{1}{2})+\frac{1}{2}\sqrt{1+4\alpha(\alpha-1)+4\ell(\ell+1)}\right]}\right]^2\frac{\hbar^2}{8\mu b^2},
\label{E31}
\end{eqnarray}
where $\zeta(\ell)=\alpha(\alpha-1)+A+\ell(\ell+1)(1-e^{1/b})$.

Using the AIM as indicated in Section $2$, we can also compute the radial eigenfuctions for the Schr\"{o}dinger equation with the Manning-Rosen potential. By comparing equation (\ref{E26}) with equation (\ref{E17}), we have the following:
\begin{eqnarray}
  a=\beta-\frac{1}{2}\ ,\ b=-1\ ,\ N=-1\ ,\ m=\gamma-\frac{1}{2}\ ,\ \sigma=2\gamma+1\ ,\ \rho=2(\gamma-\beta)+1.
  \label{E32}
\end{eqnarray}
Having determined these parameters, we can easily find the eigenfunctions $f_{n\ell}(z)$ and write the total radial wave function as
\begin{equation} R_{n\ell}(z)=N_{n\ell}(-1)^nz^\gamma(z+1)^{-\beta}\frac{\Gamma(2\alpha+n)}{\Gamma(2\alpha)}{_2F_1(-n,2(\gamma-\beta)+n+1;2\gamma+1;-z)},
\label{E33}
\end{equation}
where $N_{n\ell}$ is the normalization constant.

\section{Numerical Results}
To show the accuracy of our results, in Tables $1$ and $2$, we obtained the eigenvalues (in atomic units) numerically for arbitrary quantum numbers $n$ and $\ell$ with the potential parameter $\alpha = 0.75, 1.50$. We also computed the energy eigenvalues for HCl, CH, LiH and CO diatomic molecules as shown in Tables $3$ - $6$. It is found that the results are in good agreement with other results in the literature for short potential range as it can be seen from the results presented in these tables. The fitting parameters were obtained from \cite{VA08, IKSM10, OYKJ10, OYKJ12}.

Firstly, let us study the s-wave case $(\ell=0)$. The solutions of energy eigenvalues (\ref{E29}), (\ref{E30}) and (\ref{E31}), reduce to the following equation
\begin{equation}
E^{(Approx. 1)}_{n}=E^{(Approx. 2)}_{n}=E^{(Approx. 3)}_{n}=-\frac{\hbar^2}{2\mu b^2}\left[\frac{A-\alpha}{2(\alpha+n)}-\frac{n(n+2\alpha)}{2(\alpha+n)}\right]^2.
\label{E34}
\end{equation}
Essentialy this result coincides with the results obtained by the standard method \cite{WCQ07} when $\hbar=\mu=1$.
Secondly, we further discuss another special case with $\alpha=0,1$. As a result, the potential (\ref{E1}) becomes Hulth$\acute{e}$n potential \cite{WCQ07, SHD07, ADA05, LHA42}. So that
\begin{equation}
V(r)=-V_o\frac{e^{-\delta r}}{1-e^{-\delta r}}
\label{E35},
\end{equation}
where $V_o=\frac{A\hbar^2}{2\mu b^2}$ and $\delta=1/b$.
The corresponding energy levels for $\ell\neq0$ for this potential are given by
\begin{equation}
E^{(Approx. 1)}_{n\ell}=\left[A+\frac{\ell(\ell+1)}{12}\right]\frac{\hbar^2}{2\mu b^2}-\left[\frac{(n+\ell+1)^2+A}{n+\ell+1}\right]^2\frac{\hbar^2}{8\mu b^2},
\label{E36}
\end{equation}
\begin{eqnarray}
 E^{(Approx. 2)}_{n\ell}=\left[A+\ell(\ell+1)(1-e^\delta)\right]\frac{\hbar^2}{2\mu b^2} -\left[\frac{(n+\ell+1)^2+A+\ell(\ell+1)(1-e^\delta)}{(n+\ell+1)}\right]^2\frac{\hbar^2}{8\mu b^2},
\label{E37}
\end{eqnarray}
and
\begin{equation}
E^{(Approx. 3)}_{n\ell}=\frac{\hbar^2}{2\mu b^2}(A)-\left[\frac{(n+\ell+1)^2+A}{n+\ell+1}\right]^2\frac{\hbar^2}{8\mu b^2}.
\label{E38}
\end{equation}
Furthermore, for s-wave $(\ell=0)$ states, equations (\ref{E36}), (\ref{E37}) and (\ref{E38}) reduce to 
\begin{equation}
E^{(Approx. 1)}_{n}=E^{(Approx. 1)}_{n}=E^{(Aprox.3)}_{n}=-\frac{\left[A-(n+1)^2\right]^2}{(n+1)^2}\frac{\hbar^2}{8\mu b^2}.
\label{E39}
\end{equation}
These results are consistent with those obtained by the other methods \cite{WCQ07, SHD07}, the Feynman integral method $\cite{ADA05}$ and the N-U method $\cite{SMI11}$.

\section{Conclusions}
In this paper, we have studied analytically the solutions of the Schr\"{o}dinger equation for the Manning-Rosen potential for arbitrary $\ell$ bound states within the framework of the asymptotic iteration method. We have calculated the energy eigenvalues numerically for arbitrary $n$ and $\ell$. We have also computed the energy eigenvalues for a few HCl, Cl, LiH and CO diatomic molecules. Furthermore, we studied two special cases namely s-wave case ($\ell=0$) and Hulth$\acute{e}$n potential case ($\alpha=0,1$).  We find that the results are in good agreement with other findings in the literature, as it can be seen from the results presented in the tables.

It is evident that the AIM is efficient, accurate and an alternative method of calculating energy eigenvalues and eigenfunctions of the Manning-Rosen potential and other interaction problems that are analytically solvable. It should be emphasized that the AIM provides a closed-form for the energy eigenvalues and the corresponding eigenfunctions for exactly solvable problems. However, if there is no such a solution, the energy eigenvalues are obtained by using an iterative approach \cite{TB06, TB05, FM04}.

{\bf  \large{Acknowledgments}.}

\noindent
The authors appreciate the efforts of Profs. Hasanabadi, H., Yasuk, F. and Dr. Bahar, M. K. for their assistance. eJDS (ICTP) is acknowledged. 


\begin{table}[!hbp]
{\begin{center}
\caption{{\small Eigenvalues  (\ref{E29}), (\ref{E30}) and (\ref{E31}) (in $eV$) as a function of $\alpha$ for 2p, 3p,3d, 4p,4d, 4f, 5p, 5d, 5f, 5g, 6p, 6d, 6f and 6g states in atomic units ($\hbar=\mu=1$) and for 
$A=2b, \alpha=0.75$.}}\vspace*{2mm} {\small 
\begin{tabular}{cccccccc}\hline\hline
{}&{}&{}&{}&{}&{}&{}&{}\\[-0.5ex]
states&1/b&Approx. 1&Approx. 2&Approx. 3&LS\cite{WLF99}&QD\cite{WCQ09}&SMI\cite{SMI11}\\[1ex]\hline\hline
	 	 &0.025	&-0.120527265	&-0.120418880	&-0.120579348	&-0.1205271	&-0.1205270&-0.1205793\\[1.0ex]
2p	&0.050	&-0.108214465	&-0.107807107	&-0.108422798	&-0.1082151	&-0.1082140&-0.1084228\\[1.0ex]
   	&0.075	&-0.096443282	&-0.095588341	&-0.094099532	&-0.0964469	&-0.0964433&-0.0969120\\[1.0ex]
  	&0.025	&-0.045877611	&-0.045864389	&-0.045929694	&-0.0458779	&-0.0458776&-0.0459296\\[1.0ex]
3p	&0.050	&-0.035058868	&-0.035035701	&-0.035267202	&-0.0350633	&-0.0035058&-0.0352672\\[1.0ex]
	  &0.075	&-0.025542209	&-0.025559177	&-0.026010959	&-0.0255654	&-0.0255422&-0.0260109\\[1.0ex]
	  &0.025	&-0.044773693	&-0.044738013	&-0.044929943	&-0.0447743	&-0.0447737&-0.0449299\\[1.0ex]
3d	&0.050	&-0.033683244	&-0.033631531	&-0.034308244	&-0.0336930	&-0.0336832&-0.0343082\\[1.0ex]
	  &0.075	&-0.023710563	&-0.023808407	&-0.025116813	&-0.0237621	&-0.0237106&-0.0251168\\[1.0ex]
	  &0.025	&-0.020808737	&-0.020827991	&-0.020860820	&-0.0208097	&-0.0208087&-0.0208608\\[1.0ex]
4p	&0.050	&-0.011720853	&-0.011828800	&-0.011929186	&-0.0117365	&-0.0117209&-0.0119291\\[1.0ex]
	  &0.025	&-0.020299297	&-0.020358740	&-0.020455547	&-0.0203017	&-0.0202993&-0.0204555\\[1.0ex]
4d	&0.050	&-0.010949158	&-0.011280727	&-0.011574158	&-0.0109904	&-0.0109492&-0.0115741\\[1.0ex]
	  &0.025	&-0.019976163	&-0.020096597	&-0.020288663	&-0.0199797	&-0.0199762&-0.0202886\\[1.0ex]
4f	&0.050	&-0.010178355	&-0.010850560	&-0.011428355	&-0.0102393	&-0.0101784&-0.0114283\\[1.0ex]
5p	&0.025	&-0.009805520	&-0.009839609	&-0.009857603	&-0.0098079	&-0.0098055&-0.0098576\\[1.0ex]
5d	&0.025	&-0.009507438	&-0.009610568	&-0.009663688	&-0.0095141	&-0.0095074&-0.0096637\\[1.0ex]
5f	&0.025	&-0.009271176	&-0.009478255	&-0.009583676	&-0.0092825	&-0.0092712&-0.0095837\\[1.0ex]
5g	&0.025	&-0.009018988	&-0.009365058	&-0.009539821	&-0.0090330	&-0.0090190&-0.0095398\\[1.0ex]
6p	&0.025	&-0.004149868	&-0.004395133	&-0.004405135	&-0.0043583	&-0.0043531&-0.0044051\\[1.0ex]
6d	&0.025	&-0.003952747	&-0.004276617	&-0.004306118	&-0.0041650	&-0.0041499&-0.0043061\\[1.0ex]
6f	&0.025	&-0.003722009	&-0.004206733	&-0.004265247	&-0.0039803	&-0.0039528&-0.0042652\\[1.0ex]
6g	&0.025	&-0.003722009	&-0.004145897	&-0.004242842	&-0.0037611	&-0.0037220&-0.0042428\\[1.0ex]\hline\hline
\end{tabular}\label{tab1}}
\vspace*{1pt}
\end{center}}
\end{table}

\begin{table}[!hbp]
{\begin{center}
\caption{{\small Eigenvalues  (\ref{E29}), (\ref{E30}) and (\ref{E31}) (in $eV$) as a function of $\alpha$ for 2p, 3p,3d, 4p,4d, 4f, 5p, 5d, 5f, 5g, 6p, 6d, 6f and 6g states in atomic units ($\hbar=\mu=1$) and for 
$A=2b, \alpha=1.5$.}}\vspace*{2mm} {\small
\begin{tabular}{cccccccc}\hline\hline
{}&{}&{} &{} &{}&{} &{}&{}\\[-0.5ex]
states&1/b&Approx. 1&Approx. 2&Approx. 3&LS\cite{WLF99}&QD\cite{WCQ09}&SMI\cite{SMI11}\\[1ex]\hline\hline	 &0.025	&-0.090022888	&-0.089970805	&-0.089902617	 &-0.0899708	&-0.0899708	 &-0.0900228\\[1.0ex]	
2p	&0.050	&-0.080247211	&-0.080038878	&-0.079787754	&-0.0800400	&-0.0800389	&-0.0802472\\[1.0ex]	
	  &0.075	&-0.068220746	&-0.070564496	&-0.070050273	&-0.0705701	&-0.0705645	&-0.0710332\\[1.0ex]	
	  &0.025	&-0.036965098	&-0.036913014	&-0.036911875	&-0.0369134	&-0.0369130	&-0.0369650\\[1.0ex]	
3p	&0.050	&-0.027471932	&-0.036913014	&-0.027286328	&-0.0272696	&-0.0272636	&-0.0274719\\[1.0ex]	
	  &0.075	&-0.019385017	&-0.036913014	&-0.019030789	&-0.0189474	&-0.0189163	&-0.0193850\\[1.0ex]	
	  &0.025	&-0.039634470	&-0.039478220	&-0.039464723	&-0.0394789	&-0.0394782	&-0.0396344\\[1.0ex]	
3d	&0.050	&-0.030062923	&-0.039478220	&-0.029466438	&-0.0294496	&-0.0294379	&-0.0300629\\[1.0ex]	
	  &0.075	&-0.021812093	&-0.039478220	&-0.020664098	&-0.0204663	&-0.0204058	&-0.0218120\\[1.0ex]	
	  &0.025	&-0.017224917	&-0.017172833	&-0.017197171	&-0.0171740	&-0.0171728	&-0.0172249\\[1.0ex]	
4p	&0.050	&-0.009101874	&-0.017172833	&-0.009020329	&-0.0089134	&-0.0088935	&-0.0091019\\[1.0ex]	
	  &0.025	&-0.018364912	&-0.018208662	&-0.018277246	&-0.0182115	&-0.0182087	&-0.0183649\\[1.0ex]	
4d	&0.050	&-0.010094745	&-0.018208662	&-0.009832879	&-0.0095167	&-0.0094697	&-0.0100947\\[1.0ex]	
	  &0.025	&-0.018922274	&-0.018609774	&-0.018742772	&-0.0186137	&-0.0186098	&-0.0189222\\[1.0ex]	
4f	&0.050	&-0.010585249	&-0.018609774	&-0.010047153	&-0.0094015	&-0.0093353	&-0.0105852\\[1.0ex]	
5p	&0.025	&-0.008130783	&-0.008078700	&-0.008115365	&-0.0080816	&-0.0080787	&-0.0081308\\[1.0ex]	
5d	&0.025	&-0.008690244	&-0.008533994	&-0.008641669	&-0.0085415	&-0.0085340	&-0.0086902\\[1.0ex]	
5f	&0.025	&-0.008962206	&-0.008649706	&-0.008862907	&-0.0086619	&-0.0086497	&-0.0089622\\[1.0ex]	
5g	&0.025	&-0.009121004	&-0.008649706	&-0.008862907	&-0.0086150	&-0.0086002	&-0.0091210\\[1.0ex]	
6p	&0.025	&-0.003533421	&-0.003481337	&-0.003524889	&-0.0034876	&-0.0034813	&-0.0035334\\[1.0ex]	
6d	&0.025	&-0.003820917	&-0.003664667	&-0.003793959	&-0.0036813	&-0.0036647	&-0.0038209\\[1.0ex]	
6f	&0.025	&-0.003960631	&-0.003648131	&-0.003905471	&-0.0036774	&-0.0036481	&-0.0039606\\[1.0ex]	
6g	&0.025	&-0.004042185	&-0.003521352	&-0.003949174	&-0.0035623	&-0.0035214	&-0.0040422\\[1.0ex]\hline\hline		
\end{tabular}\label{tab2} }
\vspace*{-1pt}
\end{center}}
\end{table}

\begin{table}[!hbp]
{\begin{center}
\caption{{\small For $CH$ and $CO$,  diatomic molecules, the comparison of the energy eigenvaluess ($E_{n\ell}$) (in $eV$) by using the AIM with the other methods for 2p, 3p, 3d, 4p, 4d, 4f, 5p, 5d, 5f, 5g, 6p, 6d, 6f and 6g states with $\hbar c=1973.29 eV A^o$, $\mu_{CH}=0.929931$ $amu$, $\mu_{CO}=6.8606719$ $amu$, $\alpha=0.75$ and $A=2b$.  }}\vspace*{2pt} {\tiny
\begin{tabular}{cccccccccc}\hline\hline
\multicolumn{2}{c}{}&\multicolumn{4}{c}{$CH$$/\alpha=0.75$}&\multicolumn{4}{c}{$CO$/$\alpha=0.75$} \\[1.0ex] \hline
{}&{} &{} &{}&{}&{}&{}&{} &{}&{}\\[-1.0ex]
states&1/b&Approx. 1&Approx. 2&Approx. 3&SMI\cite{SMI11}&Approx. 1&Approx. 2&Approx. 3&SMI$\cite{SMI11}$\\[2.5ex]\hline\hline
&0.025	&-5.419968453	&-5.415094497	&-5.422310579	&-5.42025940	&-0.734650594	&-0.733989952	&-0.734968057	&-0.734690030\\[1.0ex]
2p&0.050	&-4.866276397	&-4.847957984	&-4.875644901	&-4.87380256	&-0.659600305	&-0.657117332	&-0.660870160	&-0.660620439\\[1.0ex]
&0.075	&-4.336940242	&-4.298494550	&-4.231544578	&-4.35637111	&-0.587851341	&-0.582640213	&-0.573565467	&-0.590485101\\[1.0ex]
&0.025	&-2.063061863	&-2.062467273	&-2.065403988	&-2.06461927	&-0.279638089	&-0.279557495	&-0.279955553	&-0.279849188\\[1.0ex]
3p&0.050	&-1.576555810	&-1.575514013	&-1.585924313	&-1.58532495	&-0.213694539	&-0.213553329	&-0.214964395	&-0.214883153\\[1.0ex]
&0.075	&-1.148602903	&-1.149365955	&-1.169682036	&-1.16923738	&-0.155687586	&-0.155791014	&-0.158544761	&-0.158484490\\[1.0ex]
&0.025	&-2.013419980	&-2.011815487	&-2.020446357	&-2.01968093	&-0.272909371	&-0.272691890	&-0.273861763	&-0.273758013\\[1.0ex]
3d&0.050	&-1.514695599	&-1.512370161	&-1.542801109	&-1.54221615	&-0.205309686	&-0.204994484	&-0.209119252	&-0.209039964\\[1.0ex]
&0.075	&-1.066235963	&-1.070635873	&-1.129473361	&-1.12904596	&-0.144523144	&-0.145119531	&-0.153094669	&-0.153036736\\[1.0ex]
&0.025	&-0.935744277	&-0.936610117	&-0.938086403	&-0.77458411	&-0.126835625	&-0.126952986	&-0.127153089	&-0.127104916\\[1.0ex]
4p&0.050	&-0.527072889	&-0.531927125	&-0.536441392	&-0.53623480	&-0.071442189	&-0.072100157	&-0.072712044	&-0.072684041\\[1.0ex]
&0.025	&-0.898304420	&-0.915508434	&-0.919861754	&-0.91951202	&-0.121760833	&-0.124092755	&-0.124682826	&-0.124635422\\[1.0ex]
4d&0.050	&-0.457708562	&-0.507280926	&-0.919861754	&-0.52027690	&-0.062040189	&-0.068759484	&-0.124682826	&-0.070521025\\[1.0ex]
&0.025	&-0.898304420	&-0.903720189	&-0.912357175	&-0.91200956	&-0.121760833	&-0.122494915	&-0.123665617	&-0.123618500\\[1.0ex]
4f&0.050	&-0.457708562	&-0.487936833	&-0.513919582	&-0.51372292	&-0.062040189	&-0.066137485	&-0.069659322	&-0.069632666\\[1.0ex]
5p&0.025	&-0.440942614	&-0.442475558	&-0.443284740	&-0.44311709	&-0.059767645	&-0.059975429	&-0.060085109	&-0.060062386\\[1.0ex]
5d&0.025	&-0.427538219	&-0.432175874	&-0.434564596	&-0.43440094	&-0.057950744	&-0.058579356	&-0.058903136	&-0.058880953\\[1.0ex]
5f&0.025	&-0.416913795	&-0.426225917	&-0.430966551	&-0.43080479	&-0.056510655	&-0.057772868	&-0.058415438	&-0.058393512\\[1.0ex]
5g&0.025	&-0.405573222	&-0.421135588	&-0.428994480	&-0.42883140	&-0.054973495	&-0.057082899	&-0.058148134	&-0.058126029\\[1.0ex]
6p&0.025	&-0.195751564	&-0.197643936	&-0.198093690	&-0.19801728	&-0.026533181	&-0.026789683	&-0.026850645	&-0.026840287\\[1.0ex]
6d&0.025	&-0.186614662	&-0.192314386	&-0.193641039	&-0.19356705	&-0.025294718	&-0.026067288	&-0.026247109	&-0.026237080\\[1.0ex]
6f&0.025	&-0.177750360	&-0.189171805	&-0.191803115	&-0.19172852	&-0.024093204	&-0.025641326	&-0.025997988	&-0.025987876\\[1.0ex]
6g&0.025	&-0.167374335	&-0.186436106	&-0.190795594	&-0.19072160	&-0.022686784	&-0.025270515	&-0.025861423	&-0.025851393\\[1.0ex]
\hline\hline
\end{tabular}}\label{tab3} 
\vspace*{-2pt}
\end{center}}
\end{table}

\begin{table}[!hbp]
{\begin{center}
\caption{{\small For $HCl$ and $LiH$,  diatomic molecules, the comparison of the energy eigenvaluess ($E_{n\ell}$) (in $eV$) by using the AIM with the other method for 2p, 3p, 3d, 4p, 4d, 4f, 5p, 5d, 5f, 5g, 6p, 6d, 6f and 6g states with $\hbar c=1973.29 eV A^o$, $\mu_{HCl}=0.9801045$ $amu$, $\mu_{LiH}=0.8801221$ $amu$, $\alpha=0.75$ and $A=2b$.  }}\vspace*{2pt} {\tiny
\begin{tabular}{cccccccccc}\hline\hline
\multicolumn{2}{c}{}&\multicolumn{4}{c}{$HCl$$/\alpha=0.75$}&\multicolumn{4}{c}{$LiH$/$\alpha=0.75$} \\[1.0ex] \hline
{}&{}&{} &{} &{}&{}&{} &{} &{}&{}\\[-1.0ex]
states&1/b&Approx. 1&Approx. 2&Approx. 3&SMI\cite{SMI11}&Approx. 1&Approx. 2&Approx. 3&SMI$\cite{SMI11}$\\[2.5ex]\hline\hline
  &0.025&-5.142509481&-5.137885032&-5.144731709&-5.14278553&-5.726701651&-5.721551863	&-5.729176326
&-5.72700906\\[0.8ex]
2p&0.050&-4.617162023&-4.599781366&-4.626050934&-4.62430290&-5.141674407&-5.122319296	&-5.151573103
&-5.14962650\\[0.8ex]
  &0.075&-4.114923639&-4.078446059&-4.014923390&-4.13335980&-4.582381440&-4.541759985	&-4.471021102
&-4.60291196\\[0.8ex]
	&0.025&-1.957449620&-1.956885468&-1.959671848&-1.95892730&-2.179817074&-2.179188835	&-2.182291748
&-2.18146262\\[0.8ex]
3p&0.050&-1.495848780&-1.494860315&-1.504737691&-1.50416901&-1.665778101&-1.664677346	&-1.675676798
&-1.67504351\\[0.8ex]
	&0.075&-1.089803634&-1.090527624&-1.109803684&-1.10938179&-1.213605983&-1.214412219	&-1.235878050
&-1.23540823\\[0.8ex]
	&0.025&-1.910349004&-1.908826648&-1.917015687&-1.91628944&-2.127365800&-2.125670503	&-2.134789822
&-2.13398108\\[0.8ex]
3d&0.050&-1.437155317&-1.434948922&-1.463822050&-1.46326703&-1.600417025&-1.597959983	&-1.630113115
&-1.62949505\\[0.8ex]
	&0.075&-1.011653222&-1.015827891&-1.071653372&-1.07124785&-1.126577636&-1.131226551	&-1.193393839
&-1.19294225\\[0.8ex]
	&0.025&-0.887841665&-0.888663181&-0.890063893&-0.88972668&-0.988701013&-0.989615853	&-0.991175687
&-0.99080017\\[0.8ex]
4p&0.050&-0.500090979&-0.504696716&-0.508979890&-0.50878387&-0.556901615&-0.562030567	&-0.566800311
&-0.56658202\\[0.8ex]
	&0.025&-0.872318429&-0.868641735&-0.872772200&-0.87244037&-0.979145987&-0.967319959	&-0.971919647
&-0.97155012\\[0.8ex]
4d&0.050&-0.494277550&-0.481312206&-0.872772200&-0.49364289&-0.483611741&-0.535989562	&-0.971919647
&-0.44972102\\[0.8ex]
	&0.025&-0.862429785&-0.857456954&-0.865651795&-0.86532198&-0.949142315&-0.954864579	&-0.963990360
&-0.96362308\\[0.8ex]
4f&0.050&-0.484277089&-0.462958375&-0.487611016&-0.48742442&-0.553614177&-0.515550725	&-0.543003921
&-0.54279613\\[0.8ex]
5p&0.025&-0.418369884&-0.419824353&-0.420592112&-0.42043305&-0.465896954&-0.467516652	&-0.468371628
&-0.46819450\\[0.8ex]
5d&0.025&-0.405651686&-0.410051931&-0.412318370&-0.41216309&-0.451733962&-0.456634077	&-0.459157985
&-0.45898506\\[0.8ex]
5f&0.025&-0.395571148&-0.404406564&-0.408904515&-0.40875104&-0.440508269&-0.450347393	&-0.455356314
&-0.45518540\\[0.8ex]
5g&0.025&-0.384811122&-0.399576819&-0.407033399&-0.40687867&-0.428525896&-0.444968986	&-0.453272638
&-0.45310033\\[0.8ex]
6p&0.025&-0.185730652&-0.187526149&-0.187952880&-0.18788038&-0.206829766&-0.208829233	&-0.209304440
&-0.20922370\\[0.8ex]
6d&0.025&-0.177061486&-0.182469430&-0.183728169&-0.18365796&-0.197175777&-0.203198067	&-0.204599800
&-0.20452162\\[0.8ex]
6f&0.025&-0.168650965&-0.179487724&-0.181984332&-0.18191355&-0.187809816&-0.199877637	&-0.202657861
&-0.20257904\\[0.8ex]
6g&0.025&-0.158806110&-0.176892071&-0.181028388&-0.18095818&-0.176846579&-0.196987117	&-0.201593321
&-0.20151514\\[0.8ex]\hline\hline
\end{tabular}}\label{tab4} 
\vspace*{-2pt}
\end{center}}
\end{table}

\begin{table}[!hbp]
{\begin{center}
\caption{{\small For $CH$ and $CO$,  diatomic molecules, the comparison of the energy eigenvaluess ($E_{n\ell}$) (in $eV$) by using the AIM with the other methods for 2p, 3p, 3d, 4p, 4d, 4f, 5p, 5d, 5f, 5g, 6p, 6d, 6f and 6g states with $\hbar c=1973.29 eV A^o$, $\mu_{CH}=0.929931$ $amu$, $\mu_{CO}=6.8606719$ $amu$, $\alpha=1.5$ and $A=2b$.  }}\vspace*{2pt} {\tiny
\begin{tabular}{cccccccccc}\hline\hline
\multicolumn{2}{c}{}&\multicolumn{4}{c}{$CH$$/\alpha=1.5$}&\multicolumn{4}{c}{$CO$/$\alpha=1.5$} \\[1.0ex] \hline
{}&{}&{} &{} &{}&{}&{} &{} &{}&{}\\[-1.0ex]
states&1/b&Approx. 1&Approx. 2&Approx. 3&SMI\cite{SMI11}&Approx. 1&Approx. 2&Approx. 3&SMI$\cite{SMI11}$\\[2.5ex]\hline\hline
&0.025	&-4.045880612	&-4.042814276	&-4.048222738	&-4.04668901	&-0.548399611	&-0.547983984	&-0.547983984	&-0.548509185\\[0.8ex]
2p &0.050 	&-3.599253612	&-3.587960847	&-3.608622115	&-3.60725796	&-0.487861475	&-0.486330795	&-0.657117332	&-0.488946426\\[0.8ex]
   &0.075	&-3.173201879	&-3.150077859	&-3.067806215	&-3.19307186	&-0.430112216	&-0.426977867	&-0.582640213	&-0.432805497\\[0.8ex]
 &0.025	&-1.659934569	&-1.659883336	&-1.662276695	&-1.66164415	&-0.224996128	&-0.224989184	&-0.279557495	&-0.225227854\\[0.8ex]
3p &0.050	&-1.226011781	&-1.227033867	&-1.235380284	&-1.23491200	&-0.166179986	&-0.166318525	&-0.213553329	&-0.167386368\\[0.8ex]
 &0.075	&-0.850642138	&-0.855792070	&-0.871721271	&-0.87139110	&-0.115300441	&-0.115998489	&-0.155791014	&-0.118112862\\[0.8ex]
 &0.025	&-1.775288816	&-1.774681862	&-1.782315194	&-1.78163855	&-0.240631840	&-0.240549571	&-0.272691890	&-0.241492516\\[0.8ex]
3d &0.050	&-1.323788542	&-1.325070821	&-1.351894052	&-1.35138217	&-0.179433155	&-0.179606962	&-0.204994484	&-0.183173338\\[0.8ex]
 &0.075	&-0.917626629	&-0.929240010	&-0.980864027	&-0.98048917	&-0.124379866	&-0.125954003	&-0.145119531	&-0.132900580\\[0.8ex]
 &0.025	&-0.772241983	&-0.773336440	&-0.774584109	&-0.77429066	&-0.104673678	&-0.104822026	&-0.126952986	&-0.104951366\\[0.8ex]
4p &0.050	&-0.399931993	&-0.405633524	&-0.409300496	&-0.40914700	&-0.054208853	&-0.054981669	&-0.072100157	&-0.055457903\\[0.8ex]
 &0.025	&-0.836859505	&-0.821906124	&-0.825848334	&-0.82553574	&-0.113432271	&-0.111405413	&-0.124092755	&-0.111897390\\[0.8ex]
4d &0.050	&-0.459795112	&-0.442172914	&-0.825848334	&-0.45377517	&-0.056901204	&-0.059934407	&-0.068759484	&-0.061507037\\[0.8ex]
 &0.025	&-0.867598750	&-0.842840295	&-0.850912260	&-0.85058739	&-0.113432271	&-0.114242939	&-0.122494915	&-0.115293020\\[0.8ex]
4f &0.050	&-0.477951208	&-0.451808560	&-0.476006133	&-0.47582404	&-0.061047654	&-0.061240472	&-0.066137485	&-0.064495655\\[0.8ex]
5p &0.025	&-0.363289576	&-0.364938380	&-0.365631702	&-0.36549429	&-0.049242151	&-0.049465638	&-0.059975429	&-0.049540988\\[0.8ex]
5d &0.025	&-0.383763625	&-0.388605630	&-0.390790003	&-0.39064034	&-0.052017309	&-0.05267362-	&-0.058579356	&-0.052949414\\[0.8ex]
5f &0.025	&-0.388967030	&-0.398554445	&-0.403019785	&-0.40286723	&-0.052722606	&-0.054022134	&-0.057772868	&-0.054606711\\[0.8ex]
5g &0.025	&-0.386739490	&-0.402632788	&-0.410160748	&-0.41000558	&-0.052420673	&-0.054574933	&-0.057082899	&-0.055574280\\[0.8ex]
6p &0.025	&-0.156551623	&-0.158510085	&-0.158893749	&-0.15883277	&-0.021219818	&-0.021485278	&-0.026789683	&-0.021529017\\[0.8ex]
6d &0.025 	&-0.164795756	&-0.170609838	&-0.171822133	&-0.17175642	&-0.022337270	&-0.023125341	&-0.026067288	&-0.023280755\\[0.8ex]
6f &0.025	&-0.164052130	&-0.175624409	&-0.178104885	&-0.17803620	&-0.022236475	&-0.023805042	&-0.025641326	&-0.024131947\\[0.8ex]
6g &0.025	&-0.158351033	&-0.177589686	&-0.181772291	&-0.18170426	&-0.021463719	&-0.024071426	&-0.025270515	&-0.024629136\\[0.8ex]\hline\hline
\end{tabular}}\label{tab5} 
\vspace*{-2pt}
\end{center}}
\end{table}

\begin{table}[!hbp]
{\begin{center}
\caption{{\small For $HCl$ and $LiH$,  diatomic molecules, the comparison of the energy eigenvaluess ($E_{n\ell}$) (in $eV$) by using the AIM with the other methods for 2p, 3p, 3d, 4p, 4d, 4f, 5p, 5d, 5f, 5g, 6p, 6d, 6f and 6g states with $\hbar c=1973.29 eV A^o$, $\mu_{HCl}=0.9801045$ $amu$, $\mu_{LiH}=0.8801221$ $amu$, $\alpha=1.5$ and $A=2b$.  }}\vspace*{2pt} {\tiny
\begin{tabular}{cccccccccc}\hline\hline
\multicolumn{2}{c}{}&\multicolumn{4}{c}{$HCl$$/\alpha=1.5$}&\multicolumn{4}{c}{$LiH$/$\alpha=1.5$} \\[1.0ex] \hline
{}&{}&{} &{} &{}&{}&{} &{} &{}&{}\\[-1.0ex]
states&1/b&Approx. 2&SMI\cite{SMI11}&Aprox. 2&SMI\cite{SMI11}\\[2.5ex]\hline\hline
    &0.025&-3.838763931	&-3.835854567	&-3.840986159	&-3.83953094	&-4.274849823	&-4.271609954	&-4.277324497	&-4.27570397	\\[0.8ex]
2p&0.050&-3.415000656	&-3.404285990	&-3.423889567	&-3.42259525	&-3.802946785	&-3.791014926	&-3.812845481	&-3.81140413	\\[0.8ex]
  &0.075&-3.010759359	&-2.988819104	&-2.910759110	&-3.02961216	&-3.352783434	&-3.328350752	&-3.241423095	&-3.37505550	\\[0.8ex]
&0.025&-1.574959215	&-1.574910605	&-1.577181443	&-1.57658128	&-1.753875529	&-1.753821397	&-1.756350203	&-1.75635020	\\[0.8ex]
3p&0.050&-1.163249798	&-1.164219561	&-1.172138709	&-1.17169439	&-1.295395674	&-1.296475604	&-1.305294371	&-1.30529437	\\[0.8ex]
&0.075&-0.807096074	&-0.811982371	&-0.827096124	&-0.82678285	&-0.898782674	&-0.904224057	&-0.921054742	&-0.92105474	\\[0.8ex]
&0.025&-1.684408248	&-1.683832366	&-1.691074932	&-1.69043293	&-1.875758039	&-1.875116736	&-1.883182061	&-1.88318206	\\[0.8ex]
3d&0.050&--1.257865749	&-1.257237809	&-1.282687905	&-0.93029598	&-1.400060780	&-1.400060780	&-1.428402023	&-1.42786117	\\[0.8ex]
&0.075&-0.870651496	&-0.881670364	&-0.930651645	&-0.93808640	&-0.969558029	&-0.981828648	&-1.036374232	&-1.03597816	\\[0.8ex]
&0.025&-0.732709379	&-0.733747809	&-0.734931606	&-0.73465318	&-0.815945605	&-0.817102001	&-0.818420279	&-0.81811023	\\[0.8ex]
4p&0.050&-0.379458678	&-0.384868337	&-0.388347589	&-0.38820195	&-0.422565412	&-0.428589611	&-0.432464109	&-0.43230193	\\[0.8ex]
&0.025&-0.794019001	&-0.779831114	&-0.783571514	&-0.78327492	&-0.884220038	&-0.868420398	&-0.872585709	&-0.87225543	\\[0.8ex]
4d&0.050&-0.438304965	&-0.419537203	         &-0.436756478	&-0.43054552	&-0.443552648	&-0.467196881	&-0.872585709	&-0.47945575	\\[0.8ex]
&0.025&-0.814019010	&-0.799693623	&-0.807352368	&-0.80704413	&-0.884220038	&-0.890539300	&-0.899068083	&-0.89872483	\\[0.8ex]
4f&0.050&-0.398304965	&-0.428679581	&-0.451638431	&-0.45146566	&-0.508654356	&-0.477377839	&-0.502944829	&-0.50275243	\\[0.8ex]
5p&0.025&-0.344692059	&-0.346256458	&-0.346914287	&-0.34678391	&-0.383849285	&-0.385591400	&-0.386323959	&-0.38617877	\\[0.8ex]
5d&0.025&-0.364118001	&-0.368712134	&-0.370784685	&-0.37064268	&-0.405482025	&-0.410598054	&-0.412906048	&-0.41274791	\\[0.8ex]
5f&0.025&-0.369055033	&-0.378151650	&-0.382388400	&-0.38224366	&-0.410979907	&-0.421109904	&-0.425827952	&-0.42566677	\\[0.8ex]
5g&0.025&-0.366941526	&-0.382021214	&-0.389163803	&-0.38901658	&-0.408626304	&-0.425419054	&-0.433373046	&-0.43320910	\\[0.8ex]
6p&0.025&-0.148537434	&-0.150395638	&-0.150759661	&-0.15070181	&-0.165411375	&-0.167480673	&-0.167886050	&-0.16782162	\\[0.8ex]
6d&0.025&-0.156359533	&-0.161875981	&-0.163026216	&-0.16296387	&-0.174122070	&-0.180265189	&-0.181546093	&-0.18147666	\\[0.8ex]
6f&0.025&-0.155653975	&-0.166633846	&-0.168987341	&-0.16892216	&-0.173336360	&-0.185563552	&-0.188184405	&-0.18811182	\\[0.8ex]
6g&0.025&-0.150244728	&-0.168498516	&-0.172467006	&-0.17240246	&-0.167312620	&-0.187640049	&-0.192059362	&-0.19198748	\\[0.8ex]\hline\hline
\end{tabular}}\label{tab6} 
\vspace*{-2pt}
\end{center}}
\end{table}


\begin{thebibliography}{10}
\bibitem{WCQ07} W. C. Qiang and S. H. Dong, Phys. Lett. A {\bf368} (2007) 13.
\bibitem{SHD07} S. H. Dong and J. Garcia-Ravelo, Phys. Scr. {\bf75} (2007) 307.
\bibitem{ADA05} A. Diaf, A. Chouchaoni and R. L. Lombard, Ann. Phys. (Paris) {\bf317} (2005) 354. 
\bibitem{WCQ09} W. C. Qiang, K. Li and W. L. Chen, J. Phys. A: Math. Theor. {\bf42}(2009) 205306.
\bibitem{MFM33} M. F. Manning and N. Rosen, Phys. Rev. {\bf44} (1933) 953. 
\bibitem{CBJ05} C. Berkdemir and J. Han, Chem. Phys. Lett. {\bf409} (2005) 203.
\bibitem{JPK02} J. P. Killingbeck, A. Grosjean and G. Jolicard, J. Chem. Phys. {\bf116} (2002) 447. 
\bibitem{OBI07} O. Bayrak and I. Boztosun, J. Mol. Strct. : (Theochem) {\bf802} (2007) 17.
\bibitem{MAR04} M. Akdas and  R. Sever, J. Mol. Strct. : (Theochem) {\bf710} (2004) 223.  
\bibitem{LHA42} L. Hulth$\acute{e}$n, Ark. Mat. Astron. Fys. A {\bf28} (1942) 5. 
\bibitem{BC06}  C. Berkdemir, A. Berkdemir and R. Sever, J. Phys. A: Math. Gen. {\bf39} (2006) 13455.
\bibitem{KJ10}  K. J. Oyewumi, T. T. Ibrahim, S. O. Ajibola and D. A. Ajadi, J. Vec. Rel. \textbf{5} (2010) 19.
\bibitem{KJ10B} K. J. Oyewumi and C. O. Akoshile, Eur. Phys. J. A. {\bf45} (2010) 311.
\bibitem{BJ10} 	B. J. Falaye and K. J. Oyewumi, Afr. Rev. Phys. {\bf25} (2011) 211
\bibitem{DG11} S. H. Dong, Commun. Theor. Phys. {\bf55} 969 (2011).
\bibitem{CLP34} C. L. Pekeris, Phys. Rev. {\bf45} (1934) 98.
\bibitem{RLG76} R. L. Greene and C. Aldrich, Phys. Rev. A {\bf14} (1976) 2363. 
\bibitem{SMI11} S. M. Ikhdair, ``Approximated $\ell-$states of the Manning-Rosen potential by
Nikiforov-Uvarov method" {\it arXiv:quant-ph/1110.3153v1}.
\bibitem{WLF99} W. Lucha and F. F. Sch\"{o}berl, Int. J. Mod. Phys. C {\bf10} (1999) 607. 
\bibitem{BaY12} M. K. Bahar and F. Yasuk, Few-Body Syst. DOI: 10.1007/s00601-012-0461-8.
\bibitem{HaE12} H. Hassanabadi, E. Maghsoodi, S. Zarrinkamar and H. Rahimov, Can. J. Phy. {\bf 90} (2012) 633.
\bibitem{SKN88} A. F. Nikiforov and V. B. Uvarov,  {\it Special Functions of Mathematical Physics}, (Basel, Birkhauser) (1988).
\bibitem{HCR03} H. Ciftci, R. L. Hall and N. Saad, J. Phys. A: Math Gen. {\bf36} (2003) 11807. 
\bibitem{HCR05} H. Ciftci, R. L. Hall and N. Saad, Phys. Lett. A  {\bf340} (2005) 388.
\bibitem{Fal12} B. J. Falaye, Few-Body Syst. (2012) DOI: 10.1007/s00601-012-0440-0; Cent. Eur. J. Phys. (2012) DOI: 10.24789/s11533-012-0047-6.
\bibitem{VA08}  Y. P. Varshni, Chem. Phys. {\bf 353} (1 - 3) (2008) 32.
\bibitem{IKSM10} S. M. Ikhdair, Phys. Scr. {\bf83} (2001) 015010.
\bibitem{OYKJ10} K. J. Oyewumi, F. O. Akinpelu and A. D. Agboola, Int. J. Theor. Phys. {\bf47} (2010) 1039
\bibitem{OYKJ12} K. J. Oyewumi and K. D. Sen, J. Math. Chem. {\bf50} (2012) 1039.
\bibitem{TB06} T. Barakat, J. Phys. A: Math. Gen. {\bf36} (2006) 823.
\bibitem{TB05} T. Barakat, Phys. Lett. A {\bf344} (2005) 411.
\bibitem{FM04} F. M. Fernandez, J. Phys. A: Math. Gen. {\bf37} (2004) 6173.
\end{thebibliography}
\end{document}